\documentclass[twocolumn,nopacs,preprintnumbers,amsmath,amssymb]{revtex4}

\usepackage{graphicx}
\usepackage{dcolumn}
\usepackage{bm}
\usepackage{color}

\newcommand{\ket}[1]{\left\vert{#1}\right\rangle}

\begin{document}

\title{Compact quantum circuits from one-way quantum computation}

\author{Raphael Dias da Silva}
\email{raphael.dias@if.uff.br}
\affiliation{Instituto de F\'isica, Universidade Federal Fluminense, Av. Gal. Milton Tavares de Souza s/n, 
Gragoat\'a, Niter\'oi, RJ, 24210-340, Brazil}

\author{Ernesto F. Galv\~ao}
\email{ernesto@if.uff.br}
\affiliation{Instituto de F\'isica, Universidade Federal Fluminense, Av. Gal. Milton Tavares de Souza s/n, 
Gragoat\'a, Niter\'oi, RJ, 24210-340, Brazil}

\date{\today}

\begin{abstract}
In this paper we address the problem of translating one-way quantum computation (1WQC) into the circuit model. We start by giving a straightforward circuit representation of any 1WQC, at the cost of introducing many ancilla wires. We then propose a set of four simple circuit identities that explore the relationship between the entanglement resource and correction structure of a 1WQC, allowing one to obtain equivalent circuits acting on fewer qubits. We conclude with some examples and a discussion of open problems.
\end{abstract}
\maketitle

\section{Introduction}

In the one-way model of quantum computation (1WQC) \cite{RaussendorfB01, RaussendorfB02a, RaussendorfBB01, BriegelBDRN09}, the computation is driven by one-qubit measurements on a highly entangled state, as opposed to the well-known circuit model where the information processing is driven by unitary evolution. Despite the conceptual differences, the two models were shown to be equivalent \cite{RaussendorfB02a}. In this paper we approach the problem of translating 1WQC efficiently into quantum circuits.

The one-way model requires the preparation of entangled states, and a convenient choice are the so-called cluster states \cite{RaussendorfBB03}. These states are created with the $CZ = diag(1,1,1,-1)$ unitary gate acting between qubits which are first neighbors in a regular lattice, each initially in state $\ket{+}=\frac{1}{\sqrt{2}}(\ket{0}+\ket{1})$. If the interaction happens between neighbors on more general graphs, the states created are called graph states \cite{HeinDERvdNB06}. Two-dimensional cluster states on a square lattice were shown to be universal in the 1WQC model; there has been some recent research effort to find other such universal resources \cite{GrossFE08, VanDenNestMDB6, MoraPMNDB10, WeiTAR11}.

In a realistic scenario, one may have access to a non-universal graph state and may want to know which unitaries can be implemented deterministically with it using the 1WQC model. This task has motivated the development of some methods to identify which computations can be performed via measurements on a given graph state and to describe how to implement each of the computational processes allowed. In \cite{DanosK06} a set of sufficient conditions for a given graph state to serve as a resource for implementing unitaries deterministically was found, which became known as the \textit{flow} conditions. Subsequently, a set of conditions that are both necessary and sufficient for deterministic implementation of unitaries was found in \cite{BrowneKMP07}, where they were called the \textit{gflow} (or generalized flow) conditions.

1WQC that use graphs satisfying the flow conditions have a graphical-based translation method into the circuit model called \textit{star pattern translation} \cite{DanosK06}, which results in quantum circuits having as many qubit wires as input vertices in the 1WQC graph. When one tries to apply the same method for graphs with gflow, the translation results in circuits with anachronical gates \cite{BrowneKMP07, Kashefi07}, \textit{i.e.}, two-qubit gates acting between the future and the past. Curiously, those anachronical circuits can be analyzed using models for closed timelike curves in quantum mechanics \cite{daSilvaGK11a}, being in general agreement with a closed timelike curve model that uses post-selected teleportation \cite{BennettS02, Svetlichny09}. Even though those anachronical circuits can be analyzed using these ideas, a direct translation protocol from graphs with gflow into time-respecting circuits would help in understanding the tradeoff in resources between these two models. 

The question we address here is: how can one translate a given 1WQC into a quantum circuit acting on as few qubits as possible? It is clear that any deterministic 1WQC with $n$ input/output qubits has an equivalent quantum circuit acting on $n$ qubits. Here we describe a translation method that yields such a circuit without the need for first calculating the full $n$-qubit unitary implemented by the 1WQC.

An alternative approach to this problem was proposed recently in \cite{DuncanP10}, where the use of a diagrammatic calculus allows one to rewrite a graph with gflow as a graph with flow (without changing the computation being implemented), which can then be translated into the circuit model using the star pattern translation. In this paper we provide a different translation method which does not use the star pattern translation, and yet gives the same results for graphs with flow, being also applicable to at least some graphs with gflow as well. Our method provides some insights on the origin of problems in the translation of graphs with gflow, and may hopefully lead to the development of a new, complete procedure that translates any 1WQC efficiently into the circuit model.

The paper is organized as follows. We start by reviewing the basics of the 1WQC model in Sect. \ref{sec_review}. In Sect. \ref{sec_ext} we review a straightforward method that translates any 1WQC as a circuit acting on a large number of qubits. In Sect. \ref{sec_simplifying} we propose a set of circuit identities that allow the simplification of this circuit, by removing unnecessary ancillas. We show that our method is equivalent to the star pattern translation of \cite{DanosK06} when applied to graphs with flow. Interestingly, our method can be extended to deal with some 1WQC that do not satisfy the flow conditions but do satisfy the gflow conditions. In Sect. \ref{sec_openproblems} we discuss some open problems and possible applications of our approach, and we make some concluding remarks in Sect. \ref{sec_conclusion}.

\section{1WQC review} \label{sec_review}

The one-way quantum computation model (1WQC) \cite{RaussendorfB01, RaussendorfB02a, RaussendorfBB01} can be summarized as follows. We start with a set of auxiliary qubits initialized in state $|+ \rangle \equiv \frac{1}{\sqrt{2}} (|0 \rangle + |1 \rangle)$ and a set of input qubits. We then choose some pairs of qubits to entangle with the  controlled-$Z$ gate $CZ = diag(1,1,1,-1)$, creating state $\ket{G}$ which will provide the resource for our computation. State $\ket{G}$ can be conveniently represented as a graph with qubits as nodes and $CZ$ gates represented by edges. State $|G \rangle$ has easily identifiable stabilizers, i.e. operators which leave $|G\rangle$ invariant. They are all operators of the form $K_i = X_i \prod_{j \sim i}Z_j$, where $i$ is a non-input vertex and where $j \sim i$ denotes the neighboring vertices of $i$ in the graph.

Next we would like to implement a unitary map on the input qubits, using single-qubit measurements only. It is sufficient to consider measurements onto states on the equator of the Bloch sphere. To implement the unitary in this way, the entanglement structure must be such that deterministic projections of a subset of qubits onto the chosen states implement the unitary acting on the remaining qubits. In this case, we can succeed if we are willing to post-select an exponentially small fraction of events; the 1WQC model teaches us how to obtain the same effect with unit probability, under some conditions. To understand how, we need to describe adaptive measurements.

Let $M_i^{\theta}$ represent a measurement on qubit $i$ onto basis $\{ \ket{\pm_{\theta}}\equiv 1/\sqrt{2}(\ket{0} \pm e^{i\theta}\ket{1})\}$, with outcome $s_i=0$ associated with $\ket{+_{\theta}}$, and $s_i=1$ with $\ket{-_{\theta}}$. Starting from some state $\ket{G}$, a convenient representation for the state after a deterministic projection of qubit $i$ onto state $\ket{+_{\theta}}$ is given by:
\begin{equation}
M_i^{\theta}Z^{s_i}_i\ket{G}, \label{eq_ctc}
\end{equation}
where $s_i$ is the outcome of the measurement on qubit $i$. This formal equivalence is used in \cite{BrowneKMP07}, and was given a physical justification in \cite{daSilvaGK11a}, in terms of a model for closed timelike curves in quantum mechanics. The time-ordering of the operations is from right to left. The operations in sequence (\ref{eq_ctc}) are not physically feasible, as we would need to apply $Z$ to qubit $i$ depending on the outcome $s_i$ of measurement $M_i$, which has not been performed yet. An equivalent, feasible sequence of operations can be obtained if $\ket{G}$ is stabilized by $K_j=X_j\prod_{n \sim j}Z_n, \label{eq exsta}$ as in this case we can rewrite
\begin{equation}
\ket{G}=X_j^{s_i} \prod_{n \sim j}Z_n^{s_i}\ket{G} \label{eq exsta}.
\end{equation}
If $j$ is a neighbor of $i$, there will be a cancellation of the unfeasible $Z^{s_i}_i$ operation, and we will have found a feasible sequence of operations controlled by outcomes $s_i$ which implements the same map on input qubits as sequence (\ref{eq_ctc}).

Consider the sequence of $n$ (unphysical) deterministic projections on $\ket{G}$:
\begin{equation}
M_1Z^{s_1}_1M_2Z^{s_2}_2\cdots M_nZ^{s_n}_n \ket{G}.
\end{equation}
If $\ket{G}$ has the correct set of stabilizers, each anachronical $Z^{s_j}$ operation can be eliminated, resulting in a time-respecting, feasible 1WQC whose effect on unmeasured qubits is just the same as that of the $n$ deterministic projections. The set of stabilizers used to correct the measurement on qubit $j$ is called \textit{the correcting set} for qubit $j$.

Note that rewriting state $\ket{G}$ as in Eq. (\ref{eq exsta}) may cancel anachronical $Z$ gates, but will in general result in other controlled $X$ and $Z$ gates. Each measurement $M_j$ will appear preceded by these classically controlled $Z$ and $X$ operators, which is equivalent to measuring on an adapted basis:
\begin{equation}
M_j^{\theta}Z^{t_j}X^{r_j}=M_i^{(-1)^{r_j}\theta+t_j\pi},
\end{equation}
where $t_j$ and $r_j$ are bit-valued functions which may depend on all previous measurement outcomes $\{s_i\}$. This is the adaptive nature of the measurements required to implement the 1WQC model. The correction functions $t_j$, $r_j$ for all non-input qubits $j$ also define the time-ordering required for the protocol's measurements.

Not all entanglement structures in $\ket{G}$ allow for deterministic 1WQC as described above. In \cite{DanosK06} a set of conditions called \textit{flow} was described and proven to be sufficient for deterministic implementation of unitaries with measurements on the $xy$ plane of the Bloch sphere. Later, in  \cite{BrowneKMP07} a more general set of conditions called \textit{gflow} was described and proven to be both necessary and sufficient for deterministic implementation of unitaries. The flow and gflow conditions identify when it is possible to pick a set of stabilizers that turns all unphysical deterministic projections (like the one in sequence (\ref{eq_ctc})) into physically realizable operations. For details on how to obtain measurement patterns from the flow and gflow conditions we refer the reader to references \cite{DanosK06} and \cite{BrowneKMP07}, respectively.

\begin{figure}
\includegraphics[scale=0.3]{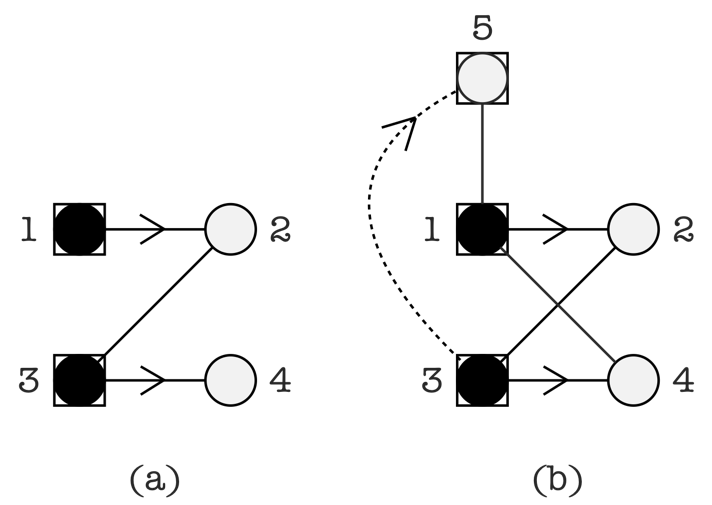}
\caption{Examples of graphs satisfying the conditions of (a) flow \cite{DanosK06} and (b) gflow \cite{BrowneKMP07}. Black vertices represent qubits to be measured, white vertices represent unmeasured qubits, and input qubits appear in a box. Arrows point from each qubit to be measured ($i$) to other vertices $j$, each of which corresponds to an operator $K_j=X_j\prod_{n \sim j}Z_n$ in the correcting set of $i$. The dashed arrow indicates a correcting stabilizer $K_j$ which is associated with a vertex $j$ not connected to $i$ in the graph.}
\label{flow_gflow}
\end{figure}

One difference between the flow and gflow conditions is that the latter corrects for measurements onto bases in the $xy, xz$ or $yz$ planes in the Bloch sphere, while in the former only the $xy$ plane is allowed. For simplicity, in this paper we will only deal with measurements in the $xy$ plane, even when dealing with graphs with gflow. Restricting measurements to the $xy$ plane still allows for 1WQC that satisfy the gflow condition but not the flow condition; the corresponding corrections are more complex than in the flow case, requiring more than one stabilizer per measurement to obtain a runnable sequence from the corresponding unphysical deterministic projection. In Fig. \ref{flow_gflow}-a we represent a graph state with flow, and in Fig. \ref{flow_gflow}-b one with gflow (but without flow). Note the graphical conventions: we use black (white) nodes to represent measured (unmeasured) qubits; and boxed nodes stand for input qubits.

\section{Extended circuit translation of 1WQC} \label{sec_ext}

In this section we review how to obtain a straightforward circuit translation from any given 1WQC, as discussed in \cite{BroadbentK09}.  This translation consists in representing in a circuit the analogue of each operation performed in the 1WQC model. As we will presently see, this method results in circuits with a large number of ancilla wires.  For this reason we call those circuits \textit{extended circuits}. These circuits will be the starting point for a more elaborated translation procedure involving the circuit identities we introduce in section \ref{sec_simplifying}.

An extended circuit is obtained as follows: for each vertex in the 1WQC graph we draw a corresponding circuit wire, which is initialized in the $|+ \rangle$ state for non-input vertices. Then, for each edge on the graph linking vertices $i$ and $j$, we draw a $CZ$ gate in the circuit acting between the wires corresponding to $i,j$. These two steps give a quantum circuit representing the entangled resource used by the 1WQC protocol. Each qubit must then be measured in a basis $\{|\pm_{\theta}\rangle \equiv \frac{1}{\sqrt{2}} (|0\rangle \pm e^{i\theta} |1\rangle)\}$; this can be represented in the circuit as a single qubit unitary followed by a measurement onto the $Z$ basis. It is easy to check that the required unitary is 
\begin{equation}
J(-\theta) = \begin{pmatrix}
 1 & e^{-i\theta} \\
 1 & -e^{-i\theta}
\end{pmatrix}.
\end{equation}

\begin{figure}
\includegraphics[scale=0.25]{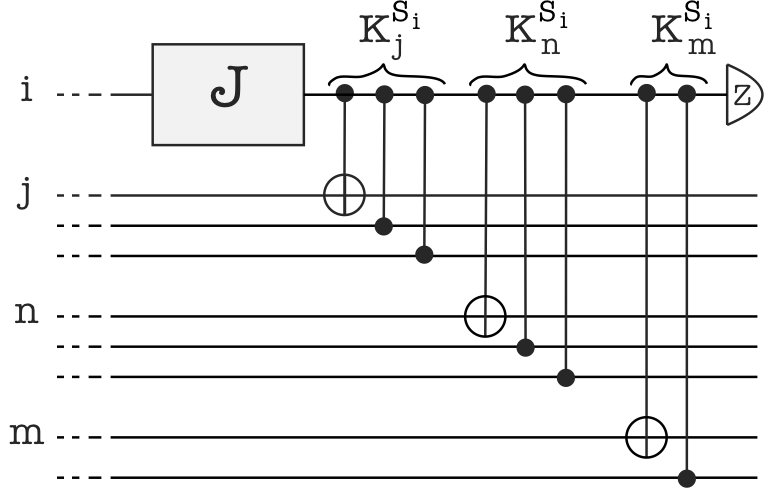}
\caption{Extended circuit representation of a measurement and the corrections it requires on other qubits for deterministic computation in the 1WQC model. Note that each correction arises from a particular stabilizer in the correcting set of $i$, raised to the power $s_i$, where $s_i$ is the outcome of measurement $i$. This sub-circuit shows only the gates corresponding to the correcting set of qubit $i$.}
\label{fig_extstab}
\end{figure}

As pointed out before, some adaptive $X$, $Z$ corrections may be needed prior to each measurement. These dependent corrections can be implemented coherently: instead of controlling the application with the classical outcome of previous measurements, we let the quantum state of each controlling previous measurement act as the control of $CX$ and $CZ$ gates. The extended circuit representation of the corrections associated with a generic measurement can be seen in Fig. \ref{fig_extstab}.

At this point it is important to note some general properties of extended circuits. They contain gates of only three types: $\{J(\theta), CZ, CX\}$. There are as many wires as there are qubits in the 1WQC graph, even though the goal is to implement a unitary on a much smaller subset of qubits. Each non-output qubit undergoes a single $J$ gate associated with its measurement in the 1WQC procedure (and wires representing output qubits have no $J$ gates). The state after each $J$ gate acts as control of possibly several $CX$ and $CZ$ gates acting on other qubit wires at a time that is after their initial entangling gates and before their respective $J$ gates.

\begin{figure}
\includegraphics[scale=0.18]{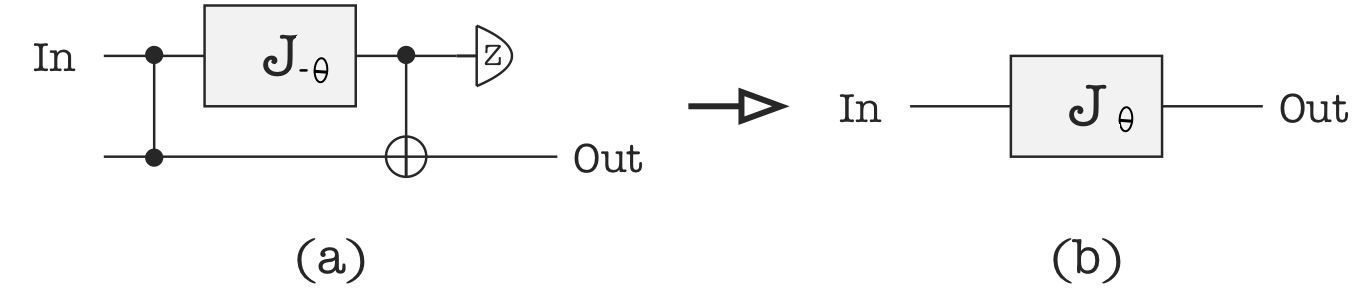}
\caption{Extended circuit for a simple one-way quantum computation protocol. This \textit{$J$-gate identity} will be used repeatedly to simplify generic extended circuits, as discussed in section \ref{sec_simplifying}. Note that the $J$ gate angles in a) and b) differ by a minus sign.}
\label{fig_jgate}
\end{figure}

As an example, let us analyze the following 1WQC sequence: $X_2^{s_1}M_1^{\theta}CZ_{12}\ket{\psi}_1\ket{+}_2$. In words, qubit 1 is an input in state $\ket{\psi}$, qubit 2 an ancilla in state $\ket{+}$; we entangle them with a $CZ$ gate, then measure qubit 1 on basis $ \{|\pm_{\theta} \rangle \equiv \frac{1}{\sqrt{2}} (|0 \rangle \pm e^{i\theta} |1 \rangle)\}$. Then we correct the state of qubit 2 by applying a Pauli $X$ conditionally on the measurement outcome $s_1$. This 1WQC sequence is represented as an extended circuit in Fig. \ref{fig_jgate}-a. Note the coherent Pauli correction: instead of a classically controlled $X$ gate, we have applied a $CX$ gate controlled by the first qubit's state. It is easy to check that the input-output map implemented by the circuits in Figures \ref{fig_jgate}-a and \ref{fig_jgate}-b is the same; we will call the identity between these circuits the \textit{$J$-gate identity}, and it will be important in the next section.

This straightforward translation into an extended circuit can be obtained for any 1WQC, as the extended circuit is just an interpretation in circuit format of the 1WQC operations, with no attempt at optimization or adaptation.

\begin{figure}
\includegraphics[scale=0.2]{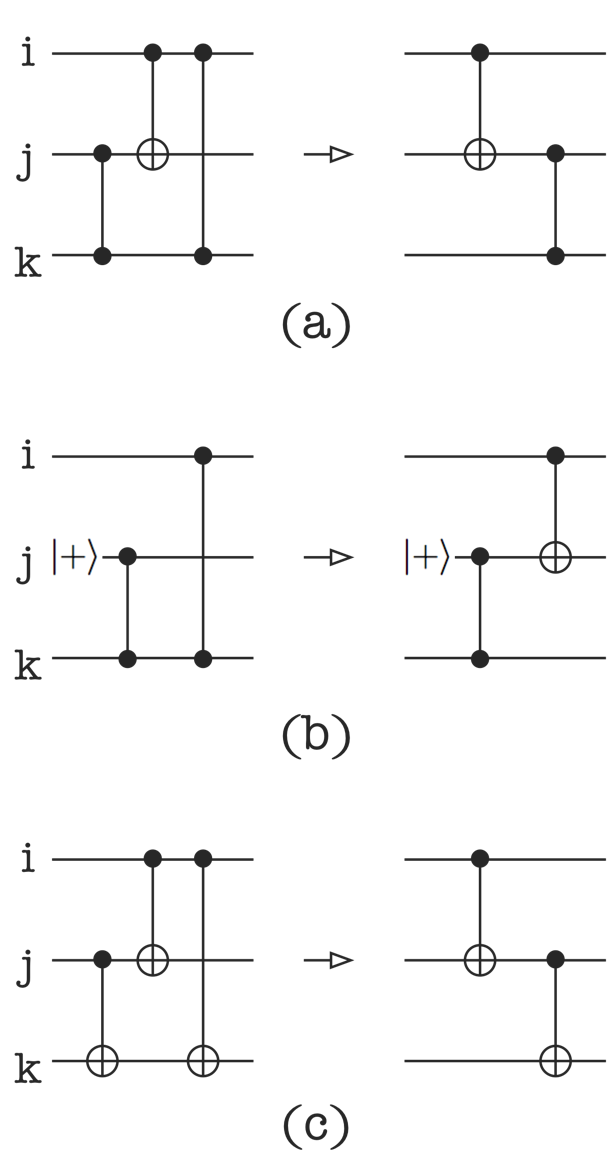}
\caption{Circuit identities that aid in the process of removing ancillas in an extended circuit corresponding to a 1WQC procedure. These circuit identities are proven in the Appendix.}
\label{fig_rules}
\end{figure}

\section{Simplifying extended circuits} \label{sec_simplifying}

In this section we propose a set of circuit identities and use them to simplify extended circuits,  obtaining equivalent circuits with a smaller number of ancilla qubits. For graphs satisfying the flow condition this translation into compact circuits is equivalent to the star pattern translation proposed in \cite{BroadbentK09}, as we will see in section \ref{sec flow}. We extend the method so that it works also for (at least some) protocols on more general graph states satisfying the gflow condition, a known limitation of the star pattern translation.

Our goal is to manipulate the extended circuit until we can repeatedly use the $J$-gate identity of Fig. \ref{fig_jgate}, which removes one ancilla wire.  Each application of the $J$-gate identity requires a few preparatory circuit manipulations. To see why, recall that corrections associated with a given measurement $M_i$ appear in the extended circuit as $CX$ and $CZ$ gates controlled by the state after the $J$ gate on qubit $i$, as in Fig. \ref{fig_extstab}. While in the circuit of Fig. \ref{fig_jgate}-a there is just a single $CX$ correction, in more general extended circuits there may be other unwanted $CX$ and $CZ$ gates that will need to be removed if we are to use the $J$ gate identity to simplify the circuit.

In Fig. \ref{fig_rules} we introduce three circuit identities that aid us in this task. The identities in Figs. \ref{fig_rules}-b and \ref{fig_rules}-c can be derived from the one in Fig. \ref{fig_rules}-a, as we describe in the Appendix. We will use the identities by substituting the subcircuit on the right for the one on the left in extended circuits, a process that does not increase the number of gates in the circuit. The identity in Fig. \ref{fig_rules}-c has been recently used in \cite{EscartinP11} for the purpose of modifying teleportation and dense coding protocols.

The  $CZ$ and $CX$ gates we would like to remove from the extended circuit using the three identities of Fig. \ref{fig_rules} are part of the correction structure of the original 1WQC. Since the correction structure differs for graphs with flow and gflow, we will discuss each case separately.

\subsection{Circuits from graphs with flow} \label{sec flow}

\begin{figure}
\includegraphics[scale=0.47]{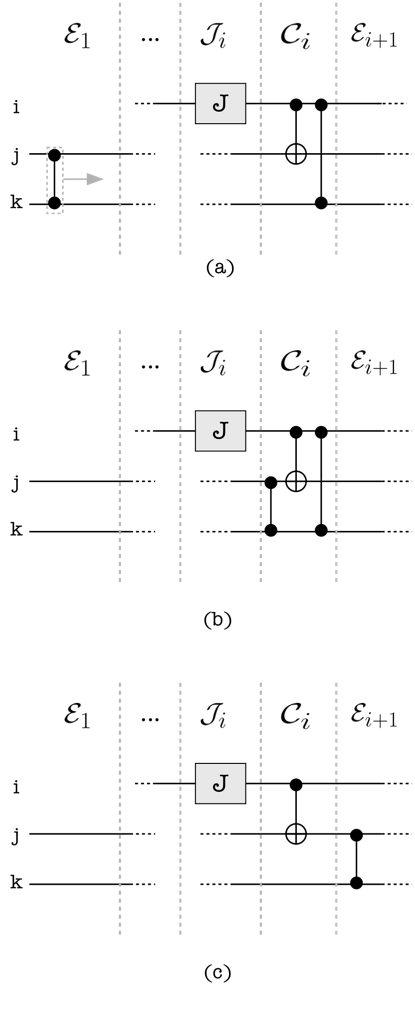}
\caption{Removing undesired $CZ$ gates. In Fig. \ref{fig_czremoval}-a we show the undesired $CZ_{ik}$ in time slice $\mathcal{C}_i$ with the corresponding, initial entangling-round $CZ_{jk}$ in time slice $\mathcal{E}_1$. In Fig. \ref{fig_czremoval}-b the $CZ_{jk}$ gate was moved from $\mathcal{E}_1$ to $\mathcal{C}_i$ where the circuit identity in Fig. \ref{fig_rules}-a can be applied, resulting in the circuit depicted in Fig. \ref{fig_czremoval}-c.}
\label{fig_czremoval}
\end{figure}

Graphs with flow have a much simpler correction structure when compared with those with gflow. Since the flow conditions require the use of just one stabilizer operator to correct each measurement, the correction structure for a given measured wire $i$ in the extended circuit will have just one $CX_{ij}$ gate, together with a set of $CZ_{ik}$ gates (for a set of controlled wires $k$). These $CZ_{ik}$ gates prevent the use of the $J$-gate identity of Fig. \ref{fig_jgate} but, as we will see, removing them is feasible. First we show how our method is able to rearrange the circuit gates, allowing subsequent applications of the $J$-gate identity. Then we show that the flow conditions ensure that our method works for any extended circuit originated from a graph with flow.

Let us start by recalling how the dependent corrections in an extended circuit are associated with the graph state's entanglement structure. As we discussed in section \ref{sec_review}, to correct for the probabilistic character of the measurement on qubit $i$ we identify operators $K_j^{s_i} = X_j^{s_i} \prod_{k \sim j} Z_k^{s_i}$, where $k$ are the neighboring vertices of a given vertex $j$ in the entanglement graph. The existence of all necessary stabilizers to attain determinism is guaranteed if either the flow or gflow conditions are met. In graphs with flow, there is a single operator $K_j^{s_i}$ needed to correct for the measurement on $i$, and the vertex $j$ associated with it is always adjacent to $i$. As in all graph states, the state's stabilizers reflect the initial entanglement resource.

As depicted in Fig. \ref{fig_extstab}, the correction operator $K_j^{s_i}$ is translated in the extended circuit as a single $CX_{ij}$ gate together with a collection of $CZ_{ik}$ gates, one for each $k\neq i$ adjacent to $j$ in the graph. There is also a $Z^{s_i}_i$ operator in $K_j^{s_i}$, but this operator does not translate as a gate in the extended circuit; instead, it cancels the anachronical $Z_i^{s_i}$ operator associated with the deterministic projection (Eq. \ref{eq_ctc}), as described previously. As the stabilizer reflects the initial graph entanglement, for each $k\neq i$ that is adjacent to $j$ in the graph there must be a $CZ_{jk}$ gate at the beginning of the extended circuit, corresponding to one of the initial entangling $CZ$ gates. 

This means that for each $CZ_{ik}$ gate we would like to remove, the extended circuit is guaranteed to have a previous $CZ_{jk}$ (from the initial entangling round of $CZ$ gates) and also a $CX_{ij}$ (that commutes with the $CZ_{ik}$ gate). These three gates can be conveniently transformed using the circuit identity in Fig. \ref{fig_rules}-a, which commutes gates $CX_{ij}$ and $CZ_{jk}$, while eliminating the troublesome $CZ_{ik}$ gate. This procedure, of bringing the corresponding $CZ$ from the beginning of the extended circuit to alongside the $CZ$ we need to remove, followed by the application of the circuit identity in Fig. \ref{fig_rules}-a, is depicted in Fig. \ref{fig_czremoval}.

Now we would like to show that this removal of unwanted $CZ$ gates can be done in extended circuits corresponding to generic graphs with flow. To see this, let us briefly review the definition of flow: Consider a graph $G$ for which we define a set $I$ of input vertices  and a set $O$ of output vertices. We define a function $f:O^c \to I^c$ (from measured to prepared qubits) and a partial order $\succ$, where $j \succ i$ means that qubit $j$ must be measured after qubit $i$. We say graph $G$ has flow if for each vertex $i \in O^c$, we can define $f$ such that:\\
(F1) $i, f(i) \in G$; \\
(F2) $f(i) \succ i$; \\
(F3) For each $k$ neighbor of $f(i)$ in $G$, with $k \neq i$, we have $k \succ i$.

When the entanglement graph has flow, $f$ is called the flow function and identifies the stabilizer $K^{s_i}_{f(i)}$ that corrects each measurement $i$, and which will be translated in the extended circuit as $CX$ and $CZ$ gates controlled by qubit $i$.

\begin{figure}
\includegraphics[scale=0.297]{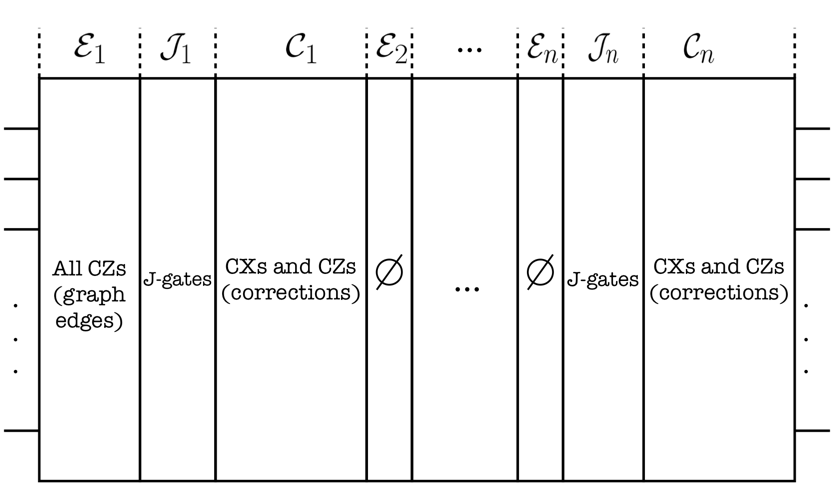}
\caption{Generic structure of extended circuits obtained from graphs with flow or gflow. See main text for information about the division into time slices.}
\label{fig_extcircsections}
\end{figure}

Flow's partial order $\succ$ defines the dependency structure of the measurement pattern, that is, it says  which qubits can be measured in a given step of the computation. Using this time ordering of gates, we can divide extended circuits into time slices which will reflect the time ordering as well as the type of gate. Let us label these time slices as $\mathcal{E}_i$, $\mathcal{J}_i$, $\mathcal{C}_i$, which respectively include the $i$th round of entangling, $J$ gates and correcting gates, as illustrated in Fig. \ref{fig_extcircsections}. For example, if the 1WQC has 3 computational steps (induced by the partial order), the corresponding extended circuit would have the following time slices: $\mathcal{E}_1\mathcal{J}_1\mathcal{C}_1\mathcal{E}_2\mathcal{J}_2\mathcal{C}_2\mathcal{E}_3\mathcal{J}_3\mathcal{C}_3$.

By construction, extended circuits are such that all $CZ$s corresponding to the entanglement structure (edges in the graph) are in slice $\mathcal{E}_1$, with slices $\mathcal{E}_2, ...,\mathcal{E}_n$ all empty. A given slice $\mathcal{J}_i$ contains the $J$ gates associated to measurements performed in the $i^{th}$ computational step of the 1WQC, and slice $\mathcal{C}_i$ contains the correcting gates ($CZ$s and $CX$s) associated to those measurements. Moreover, flow's condition (F2) implies that, for any wire $w$ such that there is a $CX$ target acting upon it in slice $\mathcal{C}_i$, the gate $J_w$ has to be in slice $J_{i+1}$ or later in order to respect the partial order $\succ$. Equivalently, flow's condition (F3) implies the same for $CZ$ in a given slice $\mathcal{C}_i$. In Fig. \ref{fig_czremoval}-a, for instance, this means that the gates $J_j$ and $J_k$ would be placed in slice $\mathcal{J}_{i+1}$ or later, since there are correcting gates acting upon wires $j$ and $k$ in slice $\mathcal{C}_i$.

Now, using the following prescription all unwanted $CZ$s can be removed. In what follows, let $W_i$ be the set of wires acted upon by the $J$ gates in slice $\mathcal{J}_i$. First, move every $CZ$ in $\mathcal{E}_1$ not acting on a wire in $W_1$ to slice $\mathcal{E}_2$, commuting with the gates in $\mathcal{C}_1$. This commutation is either trivial (no gates to be commuted) or can be done using the identity in Fig. \ref{fig_rules}-a, as done in Fig. \ref{fig_czremoval}. With this, all unwanted $CZ$ gates in $\mathcal{C}_1$ are removed. Now, since all $CZ$ gates from the entanglement structure are in $\mathcal{E}_2$ (except for those acing on wires in $W_1$), the same procedure can be repeated to remove the $CZ$ gates in $\mathcal{C}_2$, and so on until we are done. Note that, in each step, all $CZ$s needed to remove the undesired $CZ$s of a given slice $\mathcal{C}_i$ are placed in slice $\mathcal{E}_i$, which allows the application of the circuit identity in Fig. \ref{fig_rules}-a. Repeating this procedure for all slices, all unwanted $CZ$ gates can be removed.

In graphs with the same number of input and output qubits (\textit{i.e.}, $|I|=|O|$), our method is equivalent to the so-called \textit{star pattern translation} \cite{DanosK06}, which is a graphical-based approach for translating graphs with flow. To see this, first remember that an extended circuit has $|V|$ wires, where $|V|$ is the number of vertices in the associated graph. We have shown that for any flow protocol our method removes $|O^c|$ wires ($|O^c|=$ number of measured qubits) from the extended circuit, resulting in a simplified circuit with $|V|-|O^c|$ ($=|I|=|O|$) wires. We refer the reader to \cite{DanosK06} to see that the same is true for the \textit{star pattern translation} method.

Both procedures result in compact circuit translations having as many wires as the cardinality of the set of input/output qubits. One of the advantages of our procedure is that it can be extended to some graphs that do not have flow but which satisfy the gflow determinism conditions of \cite{BrowneKMP07}, as we discuss in the next section.

\subsection{Circuits from graphs with gflow} \label{sec gflow}

1WQC associated with gflow graphs may require more than one stabilizer operator to correct for each measurement. This results in more than one $X$ correction per measurement (as depicted in Fig. \ref{fig_extstab}), but in order to use our method we must identify which $CX$ gate corresponds to the one used in the $J$-gate identity of Fig. \ref{fig_jgate}, and which $CX$ gates must be removed - it is not obvious how to do this. The identification of this `special' $CX$ gate is also necessary to implement the procedure of removing $CZ$ gates (as done in Fig. \ref{fig_czremoval}), since the procedure reallocates $CZ$ gates to different wires depending on which $CX$ gate we pick to use the identity on. In what follows we will assume this `special' $CX$ gate has been identified and will concern ourselves with shifting the controls of the remaining $CX$ gates, as necessary for the use of the $J$-gate identity.

To remove these unwanted $CX$ gates we will need all three circuit identities in Fig. \ref{fig_rules}. First, note that Fig. \ref{fig_rules}-c has a very similar structure to that of Fig. \ref{fig_rules}-a, but with $CX$s instead of $CZ$s, which indicates it may be helpful in removing $CX$s in a process similar to the one we described above for $CZ$s. However, in order to use this circuit identity, another $CX$ gate is required, and it is not available in the initial entangling round, which consists of $CZ$ gates only. We can make useful $CX$ gates appear using the circuit identity in Fig. \ref{fig_rules}-b, which transforms a pair of $CZ$ gates from the initial entangling stage into one $CZ$ and one $CX$ gate.

The removal of $CX$ gates will follow basically three steps: (1) transformation of a pair of $CZ$s using the circuit identity in Fig. \ref{fig_rules}-b, which creates a new $CX$ gate; (2) moving the $CX$ originated in step (1) forward in the circuit using circuit identity \ref{fig_rules}-a when necessary, and (3) applying the circuit transformation in Fig. \ref{fig_rules}-c, where the  left-most $CX$ is the one generated in step (1), the $CX$ in the middle is the one associated to the $J$-gate identity and the right-most is the undesired $CX$, to be removed by this circuit transformation.

For the elimination of all unwanted $CX$ and $CZ$ gates, we need to identify the `special' $CX$ gate associated with each measured wire $i$; we also need to guarantee that all required initial-round $CZ$ gates are available in the extended circuit. It would be interesting to prove this is possible in general for all 1WQC graphs with gflow; we have managed to successfully apply this procedure in all small 1WQC instances we examined.

To clarify the application of this translation method and to see how more compact circuits are obtained, we now analyze a couple of examples. Note that both examples are of graphs with gflow, for which the star pattern translation method of \cite{DanosK06} fails. The dashed lines in Fig. \ref{fig_example} (first example) and Fig. \ref{fig_example2circuits} (second example) identify the set of gates that will be transformed in each step using one of the three circuit identities of Fig. \ref{fig_rules}.

\subsubsection{First example}

\begin{figure}
\includegraphics[scale=0.25]{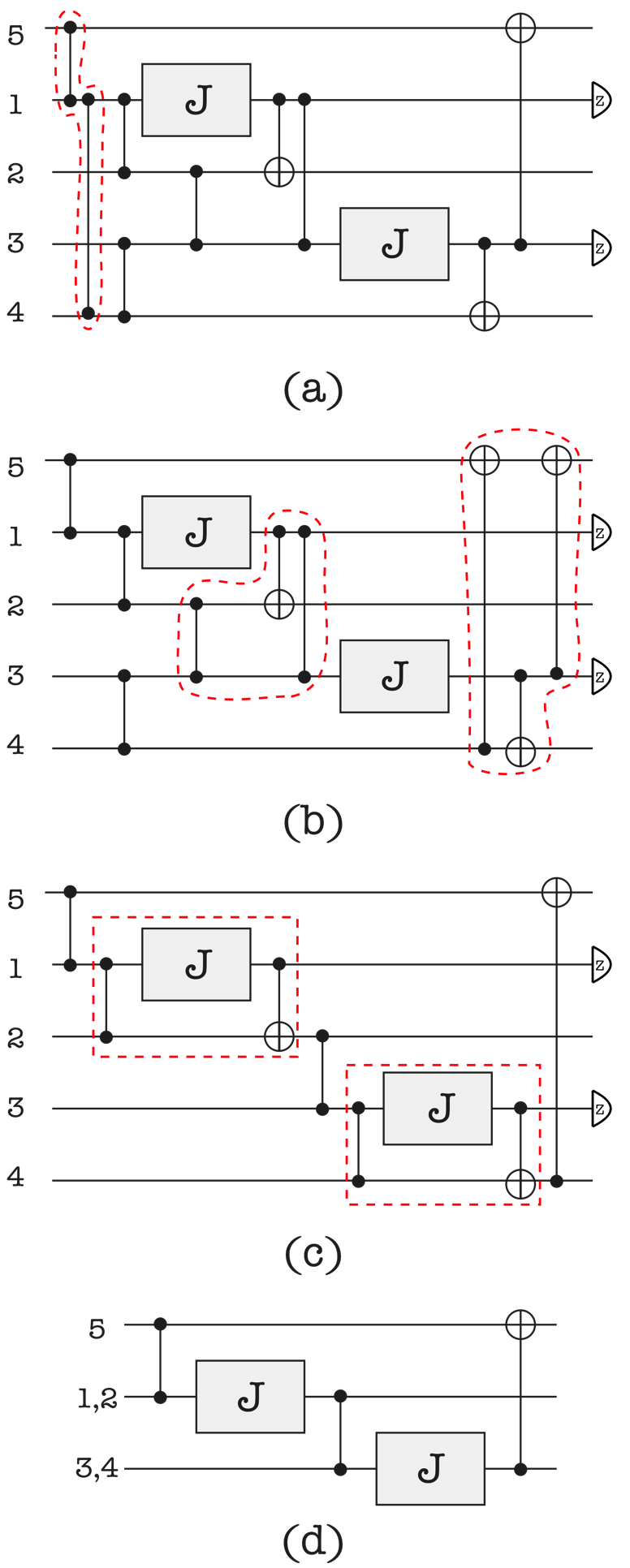}
\caption{Simplifying an extended circuit originated from the entanglement graph with gflow of Fig. \ref{flow_gflow}-b. For a step-by-step explanation, see section \ref{sec gflow}-1.}
\label{fig_example}
\end{figure}

Let us analyze the 1WQC associated with the graph with gflow in Fig. \ref{flow_gflow}-b. Since the qubit preparation and entanglement structure is already defined by the graph, we must describe the measurements and corrections. Qubits 1 and 3 are measured onto bases $\{ \ket{\pm_{\theta_i}}\equiv 1/\sqrt{2}(\ket{0} \pm e^{i\theta_i}\ket{1})\}$ with respective arbitrary angles $\theta_1$ and $\theta_3$. These measurements have correcting sets given by $g(1) = \{K_2\}$ and $g(3) = \{K_4, K_5\}$, with stabilizers $K_2 = X_2 Z_1 Z_3$, $K_4 = X_4 Z_3 Z_1$ and $K_5 = X_5 Z_1$. This information completely characterizes the 1WQC, whose associated extended circuit is shown in Fig. \ref{fig_example}-a. 

The simplification procedure for this extended circuit goes as follows. In Fig. \ref{fig_example}-a we apply the identity from Fig. \ref{fig_rules}-b, resulting in the circuit in Fig. \ref{fig_example}-b; for this circuit we need two identities: the one in Fig. \ref{fig_rules}-a for the dashed box on the left and the identity in Fig. \ref{fig_rules}-c for the box on the right. After the application of these rewrite rules, we end up with the circuit shown in Fig. \ref{fig_example}-c. We can then use the $J$-gate identity of Fig. \ref{fig_jgate} to obtain the final compact circuit in Fig. \ref{fig_example}-d.

\subsubsection{Second example}

\begin{figure}
\includegraphics[scale=0.22]{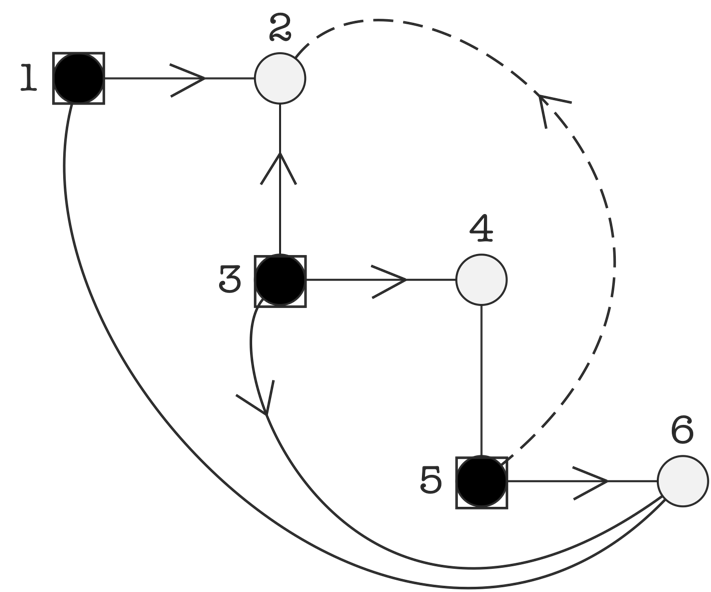}
\caption{A graph with gflow, whose extended circuit we simplify step by step in section \ref{sec gflow}-2. Qubits 1, 3, 5 are the computation's inputs, and qubits 2, 4, 6 the outputs.}
\label{fig_example2graph}
\end{figure}

In this example we will use our method to translate a more complex 1WQC example. Consider the graph in Fig. \ref{fig_example2graph}, where vertices $1$, $3$ and $5$ represent measured qubits, with respective correcting sets $g(1)=\{K_2\}$, $g(3) = \{K_2,K_4,K_6\}$ and $g(5)=\{K_2,K_6\}$. 

\begin{figure}
\includegraphics[scale=0.26]{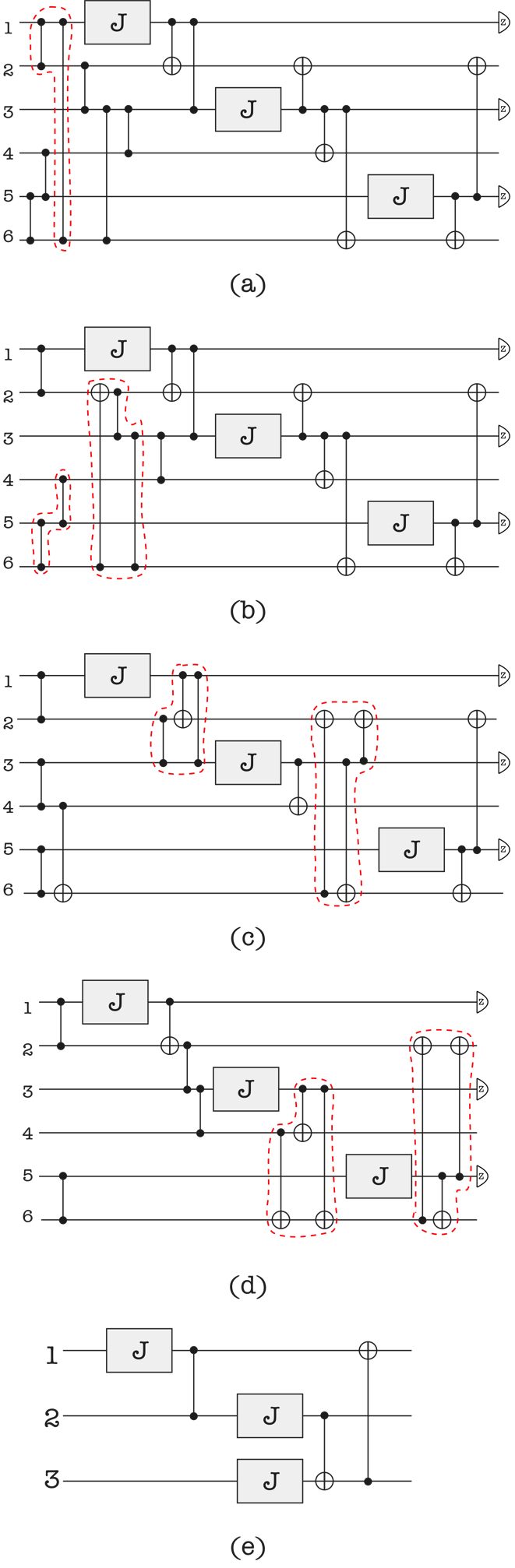}
\caption{Simplifying the extended circuit originated from the graph in Fig. \ref{fig_example2graph}. For a step-by-step explanation, see Sect. \ref{sec gflow}-2.}
\label{fig_example2circuits}
\end{figure}

In Fig. \ref{fig_example2circuits}-a the extended circuit associated with the graph in Fig. \ref{fig_example2graph} is shown, where the dashed line encloses the pair of $CZ$s that must be rewritten using the identity of Fig. \ref{fig_rules}-b. In Fig. \ref{fig_example2circuits}-b, the dashed box on the right identifies the left-hand side of the circuit identity of Fig. \ref{fig_rules}-a, which is the rule applied in this case. Also in Fig. \ref{fig_example2circuits}-b, we use the circuit identity in Fig. \ref{fig_rules}-b in the pair of $CZ$s enclosed by the box on the left. In Fig. \ref{fig_example2circuits}-c, we again use the circuit identity in Fig. \ref{fig_rules}-a to rewrite the gates in the left dashed box and use the rule in Fig. \ref{fig_rules}-c to rewrite the sequence of $CX$ gates inside the right dashed box.  The same circuit identity in Fig. \ref{fig_rules}-c is the one used to reallocate the last two undesired $CX$ gates shown in Fig. \ref{fig_example2circuits}-d.

After the application of these rewrite rules, we end up with the circuit shown in Fig. \ref{fig_example2circuits}-e, where we have already used the $J$-gate identity of Fig. \ref{fig_jgate}. Note that this final circuit has only three wires, which is the number of input vertices in the graph of Fig. \ref{fig_example2graph}.

\section{Open problems} \label{sec_openproblems}

In this section we point out some open problems regarding the generality of our translation method as well as some possible applications for it.

First let us address the method's generality issue. Since we have considered graphs with gflow with measurements only in the $xy$ plane of the Bloch sphere, the extended circuit properties discussed in section \ref{sec_ext} are particular to this scenario. The circuit identities that we proposed in this paper may be useful only for this case. One way of extending our approach would be to find a set of circuit identities that deals with the translation of 1WQC which involve measurements on the three planes:  $xy$, $xz$ and $yz$.

As we pointed out in section \ref{sec gflow}, the removal of undesired $CX$ gates (with the goal of applying the $J$-gate identity) relies on the existence of certain $CZ$ gates in the initial entangling round, as well as the identification of the `special' $CX$ gate. Although these could be identified in the examples we analyzed, a general translation protocol would require a deeper understanding of how the initial entanglement structure relates to the dependent corrections in graphs with gflow.

In \cite{BroadbentK09}, the authors studied the parallelization of quantum circuits using back-and-forth translation between 1WQC and circuit models. However, as the translation method used in that paper (the \textit{star pattern translation}) works correctly only for graphs with \textit{flow}, a full analysis including graphs with \textit{gflow} was not provided. Our translation method can be used to extend the translation for a subset of graphs with gflow. 

Interestingly, our method can also be interpreted as a way to generate graphs with flow from graphs with gflow. Using the Star Pattern Translation we can transform the circuit originated from our translation method back to 1WQC model (as done in \cite{BroadbentK09}), resulting in a graph with flow instead of gflow.

It is also interesting to look at the complexity of the simplified circuits. The rewrite rules shown in this paper simplify the correction structure of a given 1WQC, never increasing the number of gates used. It would be interesting to investigate further computational complexity tradeoffs that arise in the translation process, and our approach may help in clarifying this problem. The use of compactification procedures for  quantum circuit optimization using 1WQC techniques will be analyzed in a future publication \cite{daSilvaPKG12}.

\section{Conclusion} \label{sec_conclusion}

In this paper we have proposed a new way of translating 1WQC into quantum circuits. We start with extended circuits, which are straightforward translations of the steps in a 1WQC. These circuits necessarily involve a large number of ancillary qubits.

To the extended circuit we then apply a set of four circuit identities that explore the relationship between the entanglement and correction structures in 1WQC, allowing a reorganization of the gates in the extended circuit. This results in the removal of unnecessary wires, without increasing the number of gates in the circuit.

We have shown our method works for all graphs with flow, and also for at least some examples of graphs with gflow. We hope our work may lead the way to a new, complete and efficient translation protocol between these two very different quantum computation models.

\begin{acknowledgements}
We would like to acknowledge Elham Kashefi and Daniel Brod for helpful discussions. This work was partially funded by Brazilian agency FAPERJ and the Instituto Nacional de Ci\^{e}ncia e Tecnologia de Informa\c{c}\~{a}o Qu\^{a}ntica (INCT-IQ, CNPq). 
\end{acknowledgements}

\appendix*
\section{}

In this Appendix we show how the circuit identities in Figure \ref{fig_rules}-b and \ref{fig_rules}-c can be derived from \ref{fig_rules}-a, which can be easily verified by simple multiplication of the corresponding matrices for $CX$ and $CZ$. Note that the circuit identity in Fig. \ref{fig_rules}-a can be written as:
\begin{equation} \label{eq_rule1}
CZ_{ik}CX_{ij}CZ_{jk}=CZ_{jk}CX_{ij}
\end{equation}
with $i$, $j$ and $k$ labeling arbitrary qubits (or vertices in a graph). Note that in this representation one must read the gates from the right-hand side to the left-hand side, while in the circuit model it is the other way around. If qubit $j$ is in state $|+ \rangle$, and since $CX_{ij}|+ \rangle_j=|+ \rangle_j$, we have:
\begin{equation}
CZ_{ik}CX_{ij}CZ_{jk} |+ \rangle_j =CZ_{jk}CX_{ij}|+ \rangle_j = CZ_{jk}|+ \rangle_j 
\end{equation}
multiplying on the left by $CZ_{ik}$, omitting the $|+ \rangle_j$ and swapping the sides (to match Figure \ref{fig_rules}-b), we have:
\begin{equation}
CZ_{ik}CZ_{jk}=CX_{ij}CZ_{jk} \label{eq_rule2}
\end{equation}
which is exactly the circuit identity of of Fig. \ref{fig_rules}-b. From Eq. (\ref{eq_rule1}), considering $H_k = 1_i \otimes 1_j \otimes H_k$ and $H_k$ being the Hadamard gate, we also have:
\begin{equation}
H_k[CZ_{ik} \left ( H_k CX_{ij} H_k \right ) CZ_{jk}]H_k = H_k[CZ_{jk}]H_k CX_{ij}
\end{equation}
Since $H_kCZ_{ik}H_k=CX_{ik}$, we get:
\begin{equation} \label{eq_rule3}
CX_{ik}CX_{ij}CX_{jk}=CX_{jk}CX_{ij}
\end{equation}
which is the circuit identity in Fig. \ref{fig_rules}-c.

\end{document}